\documentclass[prb,aps,twocolumn]{revtex4-1}
\usepackage{amsmath}
\usepackage{amssymb}
\usepackage{amsthm}
\usepackage{amsfonts}
\usepackage{breqn}

\usepackage{accents}
\usepackage{listings}
\usepackage{enumerate}
\usepackage{latexsym}
\usepackage[version=4]{mhchem}

\usepackage{braket} 

\usepackage{psfrag}

\usepackage{bm}
\usepackage{graphicx}

\usepackage{color}

\usepackage{hyperref}

\makeatletter
\let\cat@comma@active\@empty
\makeatother

\catcode`_=12 \catcode`^=12 
\usepackage{tensor} 

\newcommand{\beq}{\begin{equation}}
\newcommand{\eneq}{\end{equation}}

\newcommand{\comm}[2]{\left[#1,#2\right]}
\newcommand{\anticomm}[2]{\left({#1 #2 + #2 #1}\right)}

\newcommand{\half}{\tfrac{1}{2}}
\newcommand{\ihalf}{\tfrac{i}{2}}

\renewcommand{\Re}{\,\mathrm{Re}\,}
\renewcommand{\Im}{\,\mathrm{Im}\,}

\newcommand{\expect}[1]{\left\langle {#1} \right\rangle}

\newcommand{\vary}[2]{\dfrac{\delta {#1}}{\delta {#2}}}

\newcommand{\dd}[1]{\mathrm{d}{#1}\,}

\newcommand{\inv}[1]{#1^{-1}}

\newcommand{\splus}{{\scalebox{0.5}{+}}}

\def\ie{{i.e.},\ }

\begin{document}

\tolerance 10000

\title{Hall Viscosity and the Acoustic Faraday Effect}

\author{Thomas I. Tuegel}
\author{Taylor L. Hughes}
\affiliation{Department of Physics and Institute for Condensed Matter Theory,\\
  University of Illinois at Urbana Champaign, IL 61801, USA}

\date{\today}

\begin{abstract}
For more than 20 years, observation of the non-dissipative Hall viscosity in the quantum Hall effect has been impeded by the difficulty to probe directly the momentum of the two-dimensional electron gas.
However, in three-dimensional systems such as superfluid $\ce{^3He-B}$, the momentum density is readily probed through transverse acoustic waves.
We show that in a three-dimensional elastic medium supporting transverse waves, a non-vanishing Hall viscosity induces circular birefringence.
Such an effect has been observed in $\ce{^3He-B}$ in the presence of a weak magnetic field, and is known as the acoustic Faraday effect.
The acoustic Faraday effect has been understood in terms of the Zeeman splitting of the excited order parameter modes which support the transverse wave propagation in the superfluid.
We show that the Zeeman effect can generically lead to a non-zero Hall viscosity coefficient, and confirm this prediction using a simple phenomenological model for the $\ce{^3He-B}$ collective modes.
Therefore, we claim that the observation of the acoustic Faraday effect can be leveraged to make a direct observation of the Hall viscosity in superfluid $\ce{^3He-B}$ in a magnetic field and other systems such as the crystalline $\ce{Tb3Ga5O12}$ material.
\end{abstract}

\maketitle

\section{Introduction}

Liquids with broken time-reversal symmetry can exhibit a dissipationless response to a strain rate known as the Hall viscosity.
In direct analogy to the Hall conductivity, the Hall viscosity in quantum systems may be understood as an effect of a Berry curvature of the ground state~\cite{avron95_viscos_quant_hall_fluid, lévay95_berry_landau_hamil}.
The early predictions of Hall viscosity pertained to the integer quantum Hall effect, but the Hall viscosity response is far more general,
and is present in any time-reversal breaking fluid \cite{avron98_odd_viscos}
including the fractional quantum Hall effect \cite{read09_non_abelian_hall_hall,read11_hall,abanov14_elect},
the quantum anomalous Hall effect \cite{hughes11_torsion_respon_dissip_viscos_topol_insul,hughes13_torsion_hall},
and chiral superconductors/superfluids \cite{read09_non_abelian_hall_hall,bradlyn12_kubo}.
It has also been predicted in gapless systems where strain fields couple to the low-energy electronic degrees of freedom as gauge fields (emergent elastic gauge fields)
such as graphene~\cite{cortijo15_hall_dirac},
borophene~\cite{zabolotskiy16_strain_dirac},
and Weyl semimetals~\cite{cortijo15_elast,landsteiner16_odd,liu16_circul_weyl}.

Despite the initial predictions more than two decades ago,
there are no experiments that have observed the Hall viscosity in a two-dimensional electron gas (2DEG).
The viscosity drives momentum transport, and the difficulty in directly probing the 2DEG momentum is a primary obstacle to this observation.
However, the charge current and density responses at finite wavevector~\cite{hoyos12_hall_viscos_elect_respon,nguyen14_lowes_landau_level_stres_tensor}, and the density response to spatial curvature~\cite{abanov14_elect,can14_fract_quant_hall_effec_curved_space} have been predicted to have corrections due to the Hall viscosity, and may lead to more realistic experimental proposals in the integer and fractional quantum Hall contexts.
Indeed, proposals exist to measure the Hall viscosity response
\cite{huang15_hall}
and the electron viscosity more generally
\cite{tomadin14_corbin_disk_viscom_quant_elect_liquid},
but none have yet been realized.

Interestingly, the effects of the Hall viscosity have been observed in a non-electronic system, i.e., a photonic analog of the quantum Hall effect
\cite{schine16_synth_landau}, and other
recent articles propose to measure the Hall viscosity in superfluids~\cite{fujii16_low_b_hall} and (classical) chiral active fluids
\cite{banerjee17_odd}.
Such systems may be a more expedient route to observing this response because the momentum can be probed directly.
In this article we predict another consequence of the Hall viscosity in superfluids. We claim that the Hall viscosity will generate a Faraday effect in the transverse acoustic waves in superfluid $\ce{^3He-B}$ when subjected to a magnetic field. The connection between the Hall viscosity and the acoustic Faraday effect is motivated by an experiment that has already observed this Faraday effect~\cite{lee99_discov_farad_b}.
We will focus on the connection to $\ce{^3He-B}$ here, but we note that the general concepts can be applied to other contexts where the acoustic Faraday effect has been observed or predicted, e.g., in the crystal $\ce{Tb3Ga5O12}$~\cite{strohm05_phenom,sytcheva10_acous_farad}
or in superconductor vortex lattices~\cite{dominguez95_inter,dominguez96_inter,sonin96_inter}.

Transverse acoustic waves (TAWs) provide a direct probe of momentum transport in three-dimensional paired superfluids such as $\ce{^3He-B}$, and are an unusual collective excitation for a fluid (since simple fluids usually only support longitudinal acoustic waves).
TAWs were predicted early in the history of $\ce{^3He-B}$ research \cite{landau57_oscillations,maki77_trans_b},
but were not described in detail theoretically \cite{moores93_trans_b},
or observed \cite{lee99_discov_farad_b}, until much later. This is
unlike conventional longitudinal acoustic waves that were observed much earlier
\cite{lawson73_atten_zero_sound_low_temper_trans_liquid,paulson73_propag_collis_sound_normal_extraor}.
In a typical experiment the superfluid resides in a cavity with one wall that acts as an TAW transducer. The collective mode excitations of the superfluid pairs are largely responsible for supporting the propagation of TAWs in the superfluid.
If we subject the (intrinsically time-reversal invariant) superfluid to a weak magnetic field,
the degeneracy of the collective mode excitations is broken due to the Zeeman effect.
Since the TAWs are coupled to the collective modes, the Zeeman effect, as we will see below, generates changes in the relative phase velocity of the circular polarized components of the TAWs  (i.e., circular birefringence).
For example, if a linearly-polarized TAW is injected into the fluid, then the linear-polarization rotates as it propagates through $\ce{^3He-B}$ due to the relative phase velocity of each component, a phenomenon known as the acoustic Faraday effect. We will show that this effect can be interpreted as arising directly from the Hall viscosity generated by the application of the magnetic field.

Our article is organized as follows. We begin with an overview of TAW propagation in generic fluids, and subsequently show that a Hall viscosity manifests in an acoustic Faraday effect. With this understanding, which is valid for a generic visco-elastic medium, we use the known properties of TAWs and collective modes in $\ce{^3He-B}$ to estimate the Hall viscosity coefficient at frequencies near the collective mode resonance.
Finally, we present a simple phenomenological model of the relevant $\ce{^3He-B}$ collective modes and carry out a Kubo-formula momentum response calculation to show that it exhibits the same Hall viscosity coefficient, thus corroborating our discussion.

\section{Transverse Acoustic Waves}

We begin by reviewing TAW propagation in visco-elastic media.
The conservation law for the momentum density field $g^a$ is the constitutive equation
\begin{equation}
  \partial_t g^a + \partial_b T^{ba} = 0
\end{equation}
where $T^{ab}$ is the stress tensor, and $a, b= 1, 2, 3.$
Designating the mass density by $\rho,$ and the displacement field as $u^a$, the momentum density is simply
\begin{equation}
  g^a = \rho \partial_t u^a.
\end{equation}
Throughout we will assume that the density is uniform as we are concerned with transverse, not longitudinal waves.
The stress tensor is related to the strain,
\begin{dmath}[label={eq:symm-strain}]
  w_{ab} = \half (\partial_a u_b + \partial_b u_a),
\end{dmath}
and the strain rate
\begin{math}
  \partial_t w_{ab},
\end{math}
through the elasticity $\lambda^{abcd}$ and viscosity $\eta^{abcd}$ tensors:
\begin{equation}
  T^{ab} = - \lambda^{abcd} w_{cd} - \eta^{abcd} \partial_t w_{cd}.
\end{equation}

Now let us choose a plane-wave ansatz for a TAW, i.e.,
let the displacement field be a plane wave with wavevector $\mathbf{q}$ and frequency $\omega$:
\begin{equation}
  \mathbf{u}(\mathbf{x}, t) = \mathbf{u}\, e^{i(\mathbf{q} \cdot \mathbf{x} - \omega t)}.
  \label{eq:plane-wave-displacement}
\end{equation}
In three dimensions, we can decompose the polarization vector ${\mathbf{u}}$ using an oriented triad of real, orthonormal vectors
\begin{math}
  \left\{
    \mathbf{e}_1,
    \mathbf{e}_2,
    \mathbf{e}_3
  \right\}
\end{math}
(the linear polarization basis)
such that $\mathbf{e}_3$ is chosen to be the direction of propagation:
\begin{math}
  \mathbf{q} = q \, \mathbf{e}_3.
\end{math}
We can proceed directly to the constitutive equation evaluated in this ansatz:
\begin{equation}
  \label{eq:constitutive-equation}
  - \rho \omega^2 u^a + q^2 \left( \lambda^{ac} - i \omega \eta^{ac} \right) g_{cf} u^f = 0,
\end{equation}
where, for brevity, we have designated nine components each of the elasticity and viscosity tensors as
\begin{displaymath}
  \lambda^{3a3c} = \lambda^{ac}
  \quad
  \text{and}
  \quad
  \eta^{3a3c} = \eta^{ac},\;\; a,c \in \set{1, 2, 3}.
\end{displaymath}
Note that the components with $a=3$ or $c=3$ do not contribute because the transverse waves have $u^3=0$.

In terms of classical forces,
transverse waves propagate in an elastic medium because the elasticity provides a restoring force against a strain gradient.
The ordinary viscosity appears as an imaginary term in Eq. \eqref{eq:constitutive-equation},
and generates dissipation by producing a drag force.
The anti-symmetric part of the viscosity is often neglected (e.g., in time-reversal invariant systems it vanishes),
but it leads to a force perpendicular to the strain rate gradient.
To highlight the relationship between Hall viscosity and angular momentum,
we may prefer to think of this as a non-dissipative \emph{torque}.
To see this clearly, we will work in the circular polarization basis,
\begin{math}
  \left\{
    \mathbf{e}_+,
    \mathbf{e}_-,
    \mathbf{e}_3
  \right\}
\end{math},
where
\begin{displaymath}
  \mathbf{e}_{\pm} = \tfrac{1}{\sqrt{2}} ( \mathbf{e}_1 \pm i \mathbf{e}_2 ).
\end{displaymath}
This basis has a metric given by
\begin{dgroup*}
  \begin{dmath*}
    { g_{++} = g_{--} }
    =
    {
      \mathbf{e}_+ \cdot \mathbf{e}_+
      =
      \mathbf{e}_- \cdot \mathbf{e}_-
      =
      0
    }
    \condition{and}
  \end{dmath*}
  \begin{dmath*}
    { g_{+-} = g_{-+} }
    =
    {
      \mathbf{e}_+ \cdot \mathbf{e}_-
      =
      \mathbf{e}_- \cdot \mathbf{e}_+
      =
      1
    },
  \end{dmath*}
\end{dgroup*}
so we must  take care when raising and lowering indices.
The components of a $2$-index tensor $A$ in the circular basis are related to the linear basis by
\begin{align}
  A^{++}
  = {(A^{--})}^{*}
  & = \half (A^{11} - A^{22})
    - \ihalf (A^{12} + A^{21})
  \\
  A^{+-}
  = {(A^{-+})}^{*}
  & = \half (A^{11} + A^{22})
    - \ihalf (A^{12} - A^{21})
\end{align}
where $A$ may be the (reduced) elasticity $\lambda$ or viscosity $\eta$.
The components $A^{++}$ and $A^{--}$ violate rotation symmetry
and generate coupling between right-circularly polarized waves and left-circularly polarized waves;
therefore, we demand that they vanish.
Additionally, transverse waves in $\ce{^3He-B}$ only couple to non-dissipative order parameter fluctuations
\cite{moores93_trans_b},
so we are justified to neglect the dissipative terms ($\Im\lambda^{+-}$, $\Re\eta^{+-}$).
Including the ordinary shear viscosity through the $\Re\eta^{+-}$ coefficient would introduce a contribution to the constitutive equation out of phase with the other terms, leading to a damped solution instead of an undamped plane wave.
In combination with rotation symmetry, this yields the constraints
\begin{dgroup*}
  \begin{dmath*}
    { \lambda^{12} = \lambda^{21} } = 0
  \end{dmath*}
  \begin{dmath*}
    { \eta^{11} = \eta^{22} } = 0.
  \end{dmath*}
\end{dgroup*}
The only surviving terms are the \emph{shear modulus},
\begin{dmath*}
  G = \half \left( \lambda^{11} + \lambda^{22} \right)
\end{dmath*}
and the Hall viscosity,
\begin{dmath*}
  \eta_H = \half \left( \eta^{12} - \eta^{21} \right).
\end{dmath*}

Under these conditions, the dispersion relations for the circular polarization components $\mathbf{e}_\pm$  are
\begin{dmath}[label={eq:general-dispersion}]
  \rho \omega^2 = \left( G \mp \omega \eta_H \right) q^2.
\end{dmath}
Immediately we see that each component has a different phase velocity:
the fluid exhibits circular birefringence when $\eta_H\neq 0$, \ie when time-reversal symmetry is broken.
Hence, an observation of circular birefringence in transverse acoustic waves would enable the direct determination of the Hall viscosity coefficient.
We will consider this possibility in the context of $\ce{^3He-B}$ below.

\section{Superfluid Helium 3-B}
\label{sec:superfluid-helium-3}

The superfluid $B$-phase of $\ce{^3He}$ is described by the spin-triplet, $p$-wave pairing Balian-Werthamer state
\cite{leggett75_he,halperin90_chapt_order_param_collec_modes_super,moores93_trans_b}.
The order parameter is a
\begin{math}
  3 \,\times\, 3
\end{math}
complex matrix,
\begin{equation}
  \label{eq:bw-order-param}
  d_{ij} = \frac{\Delta}{\sqrt{3}} R_{ij} (\mathbf{\hat{n}}, \theta) e^{i \phi}
\end{equation}
parameterized by the self-energy amplitude $\Delta$, the phase $\phi$, and the rotation $R_{ij}$ of the spin by an angle $\theta$ around an axis $\mathbf{\hat{n}}$ specified relative to the orbital angular momentum
\cite{moores93_trans_b}.
The ground state and collective mode excitations are eigenstates of the \emph{twisted} total angular momentum operator,
\begin{equation}
  \label{eq:twisted-angular-momentum}
  \mathbf{J} = \mathbf{L} + \inv{R} \mathbf{S}.
\end{equation}
The states are additionally labeled by their signature under particle-hole symmetry;
the real part of $d_{ij}$ transforms with signature $+1,$ and the imaginary with signature $-1$.

The six families of states
\begin{math}
  J = \set{0, 1, 2}^{\set{+, -}}
\end{math}
comprise 18 states in all.
The $J=0^-$ and $J=1^+$ states are the Goldstone modes, coupling to longitudinal zero sound and spin waves respectively
\cite{maki74_propag_balian_werth,maki76_collec_b}.
The counterparts of these modes, $J=0^+$ and $J=1^-$, are not relevant to our considerations here:
these branches of modes are at, and beyond, the pair-breaking edge and hence they are strongly damped
\cite{vollhardt90_superfluid_phases_helium}.
The real $J=2^+$ excitations do not couple to transverse waves in the quasiclassical linear response theory, so we do not consider them
\cite{mckenzie90_chapt_collec_modes_nonlin_acous_super_b,vollhardt90_superfluid_phases_helium,halperin90_chapt_order_param_collec_modes_super}.
Thus, it is only the $J=2^-$ \emph{imaginary squashing} modes that couple to TAWs below the pair-breaking frequency,
and are primarily responsible for collision-less transverse sound.

Circularly-polarized TAWs transform under the ${J = 2}$ angular momentum representation and carry ${m = \pm 1}$;
therefore, to conserve angular momentum, they couple only to the subset of the ${J = 2^-}$ multiplet with ${m = \pm 1}$~\cite{moores93_trans_b}.
The dispersion relation for TAWs with frequency $\omega$ and wavevector $q$ is given explicitly in Refs.~\onlinecite{moores93_trans_b,collett13_zeeman_split_nonlin_field_depen_super} as
\begin{dmath}[label={eq:taw-dispersion}]
  \frac{\omega^2}{v_F^2 q^2}
  = \Lambda_0
  + \Lambda_{(2^-)} \frac{\omega^2}{\omega^2 - \omega_{(2^-)}^2 - \tfrac{2}{5} v_F^2 q^2},
\end{dmath}
where, $v_F$ is the Fermi velocity, $\Lambda_0$ is the effective quasiparticle restoring force which is insensitive to the magnetic field, and 
$\Lambda_{(2^-)}$ is the coupling to the $J=2^{-}$ collective modes. 
The denominator of the second term depends on the dispersion relation of the imaginary-squashing collective modes,
\begin{dmath*}
  \omega^2 = \omega_{(2^-)}^2 + \tfrac{2}{5} v_F^2 q^2,
\end{dmath*}
where $\omega_{(2^-)}$ is the frequency edge for the $J=2^-$ modes, which will be modified by a magnetic field. In fact, the sensitivity of the TAWs to the magnetic field is primarily due to the Zeeman splitting of the $J=2^-$ collective modes.
In the following, we will show that the Zeeman effect may also be thought of as generating a contribution to the Hall viscosity, which we have already seen to be responsible for creating acoustic circular birefringence.

In zero magnetic field, the lowest energy, fully degenerate ${J=2^-}$ modes have frequency
\begin{math}
  \omega_{(2^-)} = \sqrt{\tfrac{12}{5}} \Delta.
\end{math}
We can determine the shear modulus due to the ${J=2^-}$ collective modes using Eq.~\eqref{eq:general-dispersion},
\begin{dmath*}
  \inv\rho G_{(2^-)}
  = \frac{\Lambda_{(2^-)} v_F^2 \omega^2}{\omega^2 - \tfrac{12}{5} \Delta^2 - \tfrac{2}{5} v_F^2 q^2}.
\end{dmath*}
Note that we are neglecting the quasiparticle contributions to the shear modulus because
the quasiparticle contribution is insensitive to the magnetic field (i.e., only contains time-reversal invariant contributions) and so it does not contribute to the Hall viscosity.
In the long-wavelength limit, which is valid near resonance,
\begin{dmath*}
  \lim_{q \rightarrow 0} \inv\rho G_{(2^-)}
  = \frac{\Lambda_{(2^-)} v_F^2 \omega^2}{\omega^2 - \tfrac{12}{5} \Delta^2}.
\end{dmath*}
Applying a weak magnetic field along the propagation direction breaks the $J=2^{-}$ degeneracy by Zeeman splitting:
\begin{dmath*}
  \omega_{(2^-)} =
  \sqrt{\tfrac{12}{5}} \Delta
  + m_{(2^-)} g_{(2^-)} \omega_L
\end{dmath*}
where $g_{(2^-)}$ is the Land\'e $g$-factor
\cite{schopohl81_lande_b}, $m_{(2-)}$ is the angular momentum quantum number along the applied field direction,
and $\omega_L$ is the Larmor frequency of the $J=2^-$ modes given by
\begin{dmath*}
  g_{(2^-)} \hbar \omega_L = \gamma B_z,
\end{dmath*}
where $\gamma$ is the effective coupling constant of the collective modes to the magnetic field
\cite{sauls82_inter_effec_zeeman_split_collec}.
The dispersion relation~\eqref{eq:taw-dispersion} now differs for each of the ${m_{(2^-)}=\pm 1}$ components,
\begin{dmath*}
  \frac{\omega^2}{v_F^2 q^2}
  = \Lambda_0
  + \Lambda_{(2^-)} \frac{\omega^2}{\omega^2 - {\left[ \sqrt{\tfrac{12}{5}} \Delta \pm g_{(2^-)} \omega_L \right]}^2 - \tfrac{2}{5} v_F^2 q^2}.
\end{dmath*}

Let us consider the limit near resonance in which the magnetic field $B$ is weak enough that we can consider expanding to linear order in $B$, \ie
\begin{dmath*}
  { g_{(2^-)} \omega_L \ll \sqrt{\omega^2 - \tfrac{12}{5} \Delta^2} \ll \Delta. }
\end{dmath*}
Comparing with Eq.~\eqref{eq:general-dispersion}, we find the Hall viscosity for weak magnetic fields in the long-wavelength limit to be
\begin{dmath}[label={eq:3heb-hall-viscosity}]
  \lim_{q \rightarrow 0} \eta_H
  =
  -2 G_{(2^-)} \frac{g_{(2^-)} \omega_L}{\omega^2 - \tfrac{12}{5} \Delta^2},
\end{dmath}
where we have neglected terms of order ${(\omega^2 - \tfrac{12}{5}\Delta^2)/\omega^2}$.
This shows explicitly that the Zeeman splitting of the ${J=2^-}$ collective modes may be interpreted as a direct contribution to the Hall viscosity coefficient and hence affects the TAW propagation.
We will further justify this claim by providing corroborating evidence from a Kubo formula calculation of the Hall viscosity in a phenomenological model of the collective modes. This will show the origin of the Hall viscosity from an alternate perspective.

\section{Phenomenological Model}

Let us consider a simple phenomenological model of the low-energy superfluid collective modes
as an ensemble of non-interacting bosons. We will see that this
suffices to derive a Hall viscosity coefficient that agrees with the circular birefringence result above.
The success of this model further supports our interpretation of the acoustic Faraday effect as a Zeeman-induced Hall viscosity.

We will describe an effective theory of the collective modes using a model of non-interacting bosons
with orbital angular momentum $\mathbf{\hat{L}}$ and spin $\mathbf{\hat{S}}$.
The orbital angular momentum represents, roughly, the orbital angular momentum of the quasiparticles making up the superfluid pairs.
At the mean field level, the effective interaction--experienced by the quasiparticles as they orbit--is perturbed by the strain on the system;
therefore, the orbital angular momentum will be coupled to strain.
However, the spin--an internal degree of freedom--is not coupled to the strain in this context.
Indeed, the Cooper pair has a spatial extent, so its orbital angular momentum is sensitive to the (effective) spatial metric induced by the strain.
Spin, on the other hand, is the intrinsic angular momentum of a point-like particle and insensitive to the effective strain metric.

With this assumption the dynamics of the collective modes are given by the model Hamiltonian
\begin{dmath*}
  H
  =
  H_0(\mathbf{\hat{J}})
  + \gamma \mathbf{B} \cdot \mathbf{\hat{J}}
  + \inv\mu w^{mn} \hat{L}_m \hat{L}_n
  + 2 \gamma w^{mn} B_m \hat{L}_n,
\end{dmath*}
where $\mu$ is the effective pair moment of inertia, and $\gamma$ is the effective pair coupling to the magnetic field.
The symmetrized strain tensor $w_{mn}$ is defined in Eq.~\eqref{eq:symm-strain}.
The Hamiltonian $H_0(\mathbf{\hat{J}})$ is the zero-field, zero-strain Hamiltonian for the bosons.
With $\mathbf{\hat{J}}$ given by Eq.~\eqref{eq:twisted-angular-momentum},
and the knowledge that the low-energy collective modes of $\ce{^3He-B}$ arise from the ${L=S=1}$ pairs,
it suffices for our purposes to take
\begin{dmath*}
  H_0(\mathbf{\hat{J}}) = \sum_{J=0}^2 \sum_{m=-J}^{J} \hbar \omega_{(J^-)} \ket{J, m} \bra{J, m}
\end{dmath*}
with the well-established spectrum of particle-hole antisymmetric collective modes,
\begin{displaymath}
  {\omega_{(0^-)} = 0}
  \text{, }
  {\omega_{(1^-)} = 2 \Delta}
  \text{, and }
  {\omega_{(2^-)} = \sqrt{\tfrac{12}{5}} \Delta}.
\end{displaymath}
However, we could easily have chosen $H_0$ to model another system, or to also include the particle-hole symmetric modes, and it should apply in more general contexts.

The strain susceptibility is given by the Kubo formula
\cite{kubo57_statis_mechan_theor_irrev_proces,bradlyn12_kubo}
\begin{dmath*}
  \tensor{\chi}{^{abmn}}
  = - \frac{i}{\omega} \frac{1}{\sqrt{g}}
  \expect{\vary{\tensor{T}{^{ab}}}{\tensor{w}{_{mn}}}}
  + \lim_{\epsilon \rightarrow 0^\splus} \frac{1}{\hbar\omega^\splus}
  \int_0^\infty \dd{t} e^{i \omega^\splus t}
  \expect{\comm{\tensor{T}{^{ab}}(t)}{\tensor{T}{^{mn}}}}
\end{dmath*}
where the symmetric stress tensor is
\begin{dmath*}
  \tensor{T}{^{ab}}
  =
  - \frac{1}{\sqrt{g}} \vary{H}{\tensor{w}{_{ab}}}
  =
  \frac{1}{\sqrt{g}}
  \left(
    \tensor{g}{^{am}} \tensor{g}{^{bn}}
    + \tensor{g}{^{an}} \tensor{g}{^{bm}}
  \right)
  \left[
    \frac{1}{2 \mu} \hat{L}_m \hat{L}_n
    + \gamma B_m \hat{L}_n
  \right],
\end{dmath*}
and the effective metric is
\begin{math}
  \tensor{g}{_{mn}} = \tensor{\delta}{_{mn}} + 2 \tensor{w}{_{mn}}.
\end{math}
The stress tensor is independent of $H_0$ since it is insensitive to strain.
The variational term of the susceptibility gives the infinite-frequency (contact) contribution to the transport coefficients;
if the zero-field ground state is isotropic in space,
then it gives no contribution to $\eta_H$.
The commutator term gives the finite-frequency contribution;
it vanishes unless the orbital angular momentum $L>0$ and time-reversal symmetry is broken.
This calculation is described in detail in the Appendix,
but the results are given below.

For the specific case of $\ce{^3He-B}$, the shear modulus is
\begin{dmath*}
  G_{(2^-)}
  =
  - \half i \omega \left(
    \tensor{\chi}{^1^3^1^3} + \tensor{\chi}{^2^3^2^3}
  \right)
  =
  \frac{1}{3} \frac{n}{\hbar} \left[ \frac{\hbar^2}{2 \mu} \right]^2
  \sqrt{\frac{12}{5}} \frac{\Delta}{\left(\omega^2 - \tfrac{12}{5} \Delta^2\right)}
  +
  \frac{4}{3} \frac{\hbar^2}{2 \mu} n,
\end{dmath*}
where $n$ is the boson number density.
The Hall viscosity coefficient is
\begin{dmath*}
  \eta_H
  =
  \half \left(\tensor{\chi}{^1^3^2^3} - \tensor{\chi}{^2^3^1^3}\right)
  =
  - \frac{2}{3} \frac{n}{\hbar} \left[ \frac{\hbar^2}{2 \mu} \right]^2
  \sqrt{\frac{12}{5}} \frac{\Delta}{\left(\omega^2 - \tfrac{12}{5} \Delta^2\right)^2}
  \frac{\gamma B_z}{\hbar}
\end{dmath*}
where
\begin{math}
  {\gamma B_z / \hbar = g_{(2^-)} \omega_L}.
\end{math}
Near resonance, \ie when
\begin{math}
  {\omega^2 - \tfrac{12}{5} \Delta^2 \ll \Delta^2},
\end{math}
the shear modulus is entirely dominated by its divergent term
and the Hall viscosity coefficient generated by Zeeman splitting agrees with the prediction from the previous section,
\begin{dmath*}
  \eta_H
  =
  -2 G_{(2^-)} \frac{g_{(2^-)} \omega_L}{\omega^2 - \tfrac{12}{5} \Delta^2}.
\end{dmath*}
We conclude that the Zeeman-split collective modes induce an effective Hall viscosity, which is responsible for the acoustic Faraday effect in $\ce{^3He-B}$.

\section{Discussion and Conclusion}

From our results, it may be possible to directly observe the Hall viscosity in future experiments. Several studies have been undertaken that demonstrate the acoustic Faraday effect in $\ce{^3He-B}$ using acoustic cavity interferometry.
In this type of experiment, one wall of the cavity serves as a transducer that generates and detects TAWs with a particular linear polarization.
In the low-attenuation regime, waves reflected from the opposite wall of the cavity interfere with the waves emitted by the transducer to modify the acoustic impedance.
The relative phase of the emitted and reflected waves depends on the dimensions of the cavity and on the phase velocity of the waves, which is controlled through the temperature.
As the temperature varies, the acoustic impedance oscillates in response to the changing TAW wavelength.
We have seen that, in a magnetic field, the Hall viscosity induces a relative phase velocity between polarization components, and
the relative phase velocity causes the polarization of the wave to rotate as it traverses the cavity.
Rotation of the reflected TAW away from the polarization axis of the transducer reduces the detected interference;
if the polarization of the reflected wave has rotated $90^\circ$ when it returns to the transducer, no change in the acoustic impedance is detected.
The vanishing impedance oscillations have been used to confirm the presence of the acoustic Faraday effect
\cite{lee99_discov_farad_b}
and acoustic cavity interferometry has been used further to determine precise values of the Land\'e $g$-factor
\cite{davis08_magnet_spect_super_b,collett13_zeeman_split_nonlin_field_depen_super}.
If one can modify such an experiment to collect data on the relative phase velocities then it would be possible to directly extract the Hall viscosity.

We would also like to remark briefly on the observation of the acoustic Faraday effect in other systems.
Several magneto-acoustic phenomena have been observed in $\ce{Tb3Ga5O12}$~\cite{strohm05_phenom,sytcheva10_acous_farad}.
In this crystal, a magnetic field breaks the degeneracy of phonon modes through interaction with the $\ce{Tb^{3+}}$ ions~\cite{sytcheva10_acous_farad}.
Such degeneracy breaking is reminiscent of the collective mode Zeeman effect in $\ce{^3He-B}$,
which suggests this system for future studies of the Hall viscosity.
The acoustic Faraday effect is also predicted in vortex lattices in type-II superconductors~\cite{dominguez95_inter,dominguez96_inter,sonin96_inter},
where the Magnus force between vortices enters the acoustic wave dispersion relation as a Hall viscosity term.
We are not aware of any experiment observing this effect, but the similarity to the phenomenon described here indicates it as a promising avenue of investigation.

\bibliography{hall-viscosity-acoustic-faraday}

\appendix*
\section{Calculation of the Kubo Formula for Hall Viscosity}

To calculate the Hall viscosity near resonance,
we treat the collective mode excitations as a non-interacting gas of bosons
with dynamics given by the Hamiltonian
\begin{dmath*}
  H
  =
  H_0(\mathbf{\hat{J}})
  + \gamma \mathbf{B} \cdot \mathbf{\hat{J}}
  + \inv\mu w^{mn} \hat{L}_m \hat{L}_n
  + 2 \gamma w^{mn} B_m \hat{L}_n.
\end{dmath*}
Our interpretation of this Hamiltonian in terms of the superfluid collective modes is given in the main text.
We remind the reader of the following facts:
\begin{itemize}
\item The Hamiltonian acts on the Hilbert space of eigenstates of $J^2$ and $J_3$ where
  \begin{dmath*}
    \mathbf{\hat{J}} = \mathbf{\hat{L}} + \inv{R} \mathbf{\hat{S}}
  \end{dmath*}
  and $L=S=1$ so that $J\in\set{0,1,2}$.
  These represent the particle-hole antisymmetric collective modes of $\ce{^3He-B}$.
\item At zero strain in zero field, we model the collective modes with the Hamiltonian
  \begin{dmath*}
    { H_0(\mathbf{\hat{J}}) = } \sum_{J=0}^2 \sum_{m=-J}^{J} \hbar \omega_{(J^-)} \ket{J, m} \bra{J, m}.
  \end{dmath*}
  with the collective mode spectrum
  $\omega_{(0^-)}=0$,
  $\omega_{(1^-)}=2\Delta$, and
  $\omega_{(2^-)}=\sqrt{\tfrac{12}{5}}\Delta$.
\end{itemize}
The stress susceptibility is given by the Kubo formula
\begin{dmath*}
  \tensor{\chi}{^{abmn}}
  = - \frac{i}{\omega} \frac{1}{\sqrt{g}}
  \expect{\vary{\tensor{T}{^{ab}}}{\tensor{w}{_{mn}}}}
  + \lim_{\epsilon \rightarrow 0^\splus} \frac{1}{\hbar\omega^\splus}
  \int_0^\infty \dd{t} e^{i \omega^\splus t}
  \expect{\comm{\tensor{T}{^{ab}}(t)}{\tensor{T}{^{mn}}}}.
\end{dmath*}
The symmetric stress tensor is
\begin{dmath*}
  \tensor{T}{^{ab}}
  =
  - \frac{1}{\sqrt{g}} \vary{H}{\tensor{w}{_{ab}}}
  =
  \frac{1}{\sqrt{g}}
  \left(
    \tensor{g}{^{am}} \tensor{g}{^{bn}}
    + \tensor{g}{^{an}} \tensor{g}{^{bm}}
  \right)
  \left[
    \frac{1}{2 \mu} \hat{L}_m \hat{L}_n
    + \gamma B_m \hat{L}_n
  \right]
\end{dmath*}
in terms of the effective metric
\begin{math}
  \tensor{g}{_{mn}} = \tensor{\delta}{_{mn}} + 2 \tensor{w}{_{mn}}.
\end{math}
At zero strain, this yields
\begin{dmath*}
  \left. \tensor{T}{^a^b} \right|_{w = 0}
  =
  - \dfrac{1}{2\mu} \anticomm{\hat{L}^a}{\hat{L}^b}
  - \gamma \left( B^a \hat{L}^b + B^b \hat{L}^a \right).
\end{dmath*}
The instantaneous term in the susceptibility is the expectation value at zero strain of
\begin{widetext}
\begin{dmath*}
  \left. \frac{1}{\sqrt{g}}
    \vary{\tensor{T}{^a^b}}{\tensor{w}{_m_n}}
  \right|_{w=0}
  =
  - \tensor{T}{^a^b} \tensor{\delta}{^m^n}
  - \left(
    \tensor{\delta}{^m^a} \left[
      \dfrac{1}{2\mu} \anticomm{\hat{L}^n}{\hat{L}^b}
      + \gamma (B^n \hat{L}^b + B^b \hat{L}^n)
    \right]
    + \left[ {m \leftrightarrow n} \right]
    + \left[ {a \leftrightarrow b} \right]
    + \left[ {m a \leftrightarrow n b} \right]
  \right).
\end{dmath*}
\end{widetext}
The Kubo formula enables us to calculate the following response coefficients defined in the text:
\begin{dgroup*}
  \begin{dmath*}
    G_{(2^-)}
    =
    \frac{1}{3} \frac{n}{\hbar} \left[ \frac{\hbar^2}{2 \mu} \right]^2
    \sqrt{\frac{12}{5}} \frac{\Delta}{\left(\omega^2 - \tfrac{12}{5} \Delta^2\right)}
    +
    \frac{4}{3} \frac{\hbar^2}{2 \mu} n,
  \end{dmath*}
  \begin{dmath*}
    \eta_H
    =
    - \frac{2}{3} \frac{n}{\hbar} \left[ \frac{\hbar^2}{2 \mu} \right]^2
    \sqrt{\frac{12}{5}} \frac{\Delta}{\left(\omega^2 - \tfrac{12}{5} \Delta^2\right)^2}
    \frac{\gamma B_z}{\hbar}
  \end{dmath*}
\end{dgroup*}
where $n$ is the boson number density and
\begin{math}
  \gamma B_z / \hbar = g_{(2^-)} \omega_L.
\end{math}
Near resonance, the shear modulus is entirely dominated by its first term,
\begin{dseries*}
  \begin{math}
    { G_{(2^-)} = }
    \frac{1}{3} \frac{n}{\hbar} \left[ \frac{\hbar^2}{2 \mu} \right]^2
    \sqrt{\frac{12}{5}} \frac{\Delta}{\left(\omega^2 - \tfrac{12}{5} \Delta^2\right)}
  \end{math}
  when
  \begin{math}
    \omega^2 - \tfrac{12}{5} \Delta^2 \ll \Delta^2,
  \end{math}
\end{dseries*}
and the Hall viscosity is, in terms of the shear modulus,
\begin{dmath*}
  \eta_H
  =
  -2 G_{(2^-)} \frac{g_{(2^-)} \omega_L}{\omega^2 - \tfrac{12}{5} \Delta^2}.
\end{dmath*}

\end{document}